\documentclass[aps,prl,showpacs,twocolumn,superscriptaddress]{revtex4}
\usepackage{graphicx,amsmath,amsfonts}
\usepackage{longtable}

\RequirePackage{xspace}

\newcommand{\BABARPubYear}     {06}
\newcommand{\BABARPubNumber}  {038}
\newcommand{\SLACPubNumber} {12155}
\newcommand{\LANLNumber}  {0610057}

\RequirePackage{xspace}

\usepackage{relsize}
\def\babar{\mbox{\slshape B\kern-0.1em{\smaller A}\kern-0.1em
    B\kern-0.1em{\smaller A\kern-0.2em R}}}

\def\ee         {\ensuremath{e^-e^-}\xspace}

\def\ppz   {\ensuremath{\pi^0\pi^0}\xspace}

\def\pipi  {\ensuremath{\pi^+\pi^-}\xspace}

\def\Kbar  {\kern 0.2em\overline{\kern -0.2em K}{}\xspace}

\def\Kz    {\ensuremath{K^0}\xspace}
\def\Kzb   {\ensuremath{\Kbar^0}\xspace}
\def\KzKzb {\ensuremath{\Kz \kern -0.16em \Kzb}\xspace}
\def\Kp    {\ensuremath{K^+}\xspace}
\def\Km    {\ensuremath{K^-}\xspace}

\def\KpKm  {\ensuremath{\Kp \kern -0.16em \Km}\xspace}

\def\Dbar    {\kern 0.2em\overline{\kern -0.2em D}{}\xspace}

\def\Dz      {\ensuremath{D^0}\xspace}
\def\Dzb     {\ensuremath{\Dbar^0}\xspace}
\def\DzDzb   {\ensuremath{\Dz {\kern -0.16em \Dzb}}\xspace}
\def\Dp      {\ensuremath{D^+}\xspace}
\def\Dm      {\ensuremath{D^-}\xspace}

\def\DpDm    {\ensuremath{\Dp {\kern -0.16em \Dm}}\xspace}

\def\Bbar    {\kern 0.18em\overline{\kern -0.18em B}{}\xspace}

\def\Bz      {\ensuremath{B^0}\xspace}
\def\Bzb     {\ensuremath{\Bbar^0}\xspace}
\def\BzBzb   {\ensuremath{\Bz {\kern -0.16em \Bzb}}\xspace}
\def\Bu      {\ensuremath{B^+}\xspace}
\def\Bub     {\ensuremath{B^-}\xspace}

\def\BpBm    {\ensuremath{\Bu {\kern -0.16em \Bub}}\xspace}

\def\BorBbar    {\kern 0.18em\optbar{\kern -0.18em B}{}\xspace}
\def\DorDbar    {\kern 0.18em\optbar{\kern -0.18em D}{}\xspace}
\def\KorKbar    {\kern 0.18em\optbar{\kern -0.18em K}{}\xspace}

\def\jpsi     {\ensuremath{{J\mskip -3mu/\mskip -2mu\psi\mskip 2mu}}\xspace}
\def\psitwos  {\ensuremath{\psi{(2S)}}\xspace}

\mathchardef\Upsilon="7107
\def\Y#1S{\ensuremath{\Upsilon{(#1S)}}\xspace}% no space before {...}!

\mathchardef\Deltares="7101
\mathchardef\Xi="7104
\mathchardef\Lambda="7103
\mathchardef\Sigma="7106
\mathchardef\Omega="710A

\def\Deltabar{\kern 0.25em\overline{\kern -0.25em \Deltares}{}\xspace}
\def\Lbar{\kern 0.2em\overline{\kern -0.2em\Lambda\kern 0.05em}\kern-0.05em{}\xspace}
\def\Sigbar{\kern 0.2em\overline{\kern -0.2em \Sigma}{}\xspace}
\def\Xibar{\kern 0.2em\overline{\kern -0.2em \Xi}{}\xspace}
\def\Obar{\kern 0.2em\overline{\kern -0.2em \Omega}{}\xspace}
\def\Nbar{\kern 0.2em\overline{\kern -0.2em N}{}\xspace}
\def\Xb{\kern 0.2em\overline{\kern -0.2em X}{}\xspace}

\newcommand{\tev}{\ensuremath{\mathrm{\,Te\kern -0.1em V}}\xspace}
\newcommand{\gev}{\ensuremath{\mathrm{\,Ge\kern -0.1em V}}\xspace}
\newcommand{\mev}{\ensuremath{\mathrm{\,Me\kern -0.1em V}}\xspace}
\newcommand{\kev}{\ensuremath{\mathrm{\,ke\kern -0.1em V}}\xspace}
\newcommand{\ev}{\ensuremath{\mathrm{\,e\kern -0.1em V}}\xspace}
\newcommand{\gevc}{\ensuremath{{\mathrm{\,Ge\kern -0.1em V\!/}c}}\xspace}
\newcommand{\mevc}{\ensuremath{{\mathrm{\,Me\kern -0.1em V\!/}c}}\xspace}
\newcommand{\gevcc}{\ensuremath{{\mathrm{\,Ge\kern -0.1em V\!/}c^2}}\xspace}
\newcommand{\mevcc}{\ensuremath{{\mathrm{\,Me\kern -0.1em V\!/}c^2}}\xspace}
\def\mm   {\ensuremath{{\rm \,mm}}\xspace}

\def\invfb   {\ensuremath{\mbox{\,fb}^{-1}}\xspace}

\def\mus  {\ensuremath{\rm \,\mus}\xspace}

\def\mus        {\ensuremath{\,\mu{\rm s}}\xspace}    %% microsecond
\def\order{{\ensuremath{\cal O}}\xspace}

\def\to                 {\ensuremath{\rightarrow}\xspace}

\def\pep2{PEP-II}
\def\BF{$B$ Factory}

\def\gsim{{~\raise.15em\hbox{$>$}\kern-.85em
          \lower.35em\hbox{$\sim$}~}\xspace}
\def\lsim{{~\raise.15em\hbox{$<$}\kern-.85em
          \lower.35em\hbox{$\sim$}~}\xspace}

\xspace

\newcommand{\jprlBase}       {Phys.\ Rev.\ Lett.\xspace}
\newcommand{\jprBase}        {Phys.\ Rev.\xspace}
\newcommand{\jplBase}        {Phys.\ Lett.\xspace}
\newcommand{\nimBaseA}       {Nucl.\ Instr.\ Methods Phys.\ Res., Sect.\ A\xspace}

\newcommand{\nima}      [1]  {\nimBaseA~{\bf #1}}

\newcommand{\plb}       [1]  {\jplBase\ B~{\bf #1}}

\newcommand{\jprl}      [1]  {\jprlBase\ {\bf #1}}
\newcommand{\jprd}      [1]  {\jprBase\ D~{\bf #1}}
\def\jetset74   {\mbox{\tt Jetset \hspace{-0.5em}7.\hspace{-0.2em}4}\xspace}

\def\BaBar{\babar\xspace}

\def\etal{{\it et al.}}

\newcommand{\bfg}{\begin{figure}}
\newcommand{\efg}{\end{figure}}
\newcommand{\infg}{\includegraphics}  

\newcommand{\btbl}{\begin{table}}
\newcommand{\etbl}{\end{table}}
\newcommand{\btbu}{\begin{tabular}} 
\newcommand{\etbu}{\end{tabular}}  

\newcommand{\abs}[1]{\lvert#1\rvert}

\def\jsi{\jpsi}
\def\psip{\psitwos}

\def\Xstat   {\ensuremath{X(3872)}\xspace}
\def\Ystat   {\ensuremath{Y(4260)}\xspace}

\def\ee      {\ensuremath{e^+e^-}\xspace}
\def\mm      {\ensuremath{\mu^+\mu^-}\xspace}
\def\ll      {\ensuremath{\ell^+\ell^-}\xspace}
\def\pipi    {\ensuremath{\pi^+\pi^-}\xspace}

\def\pipiJ   {\ensuremath{\pipi\jsi}\xspace}
\def\fourPiJsi {\ensuremath{2(\pipi)\jsi}\xspace}

\def\smallISR{\scalebox{0.5}{ISR}}
\def\gISR    {\ensuremath{\gamma_{\smallISR}}\xspace}
\def\eff     {\ensuremath{\varepsilon}\xspace}
\def\BF      {\ensuremath{{\cal B}}\xspace}
\def\Lumi    {\ensuremath{{\cal L}}\xspace}

\def\DelP    {\ensuremath{\Delta p^{*}}\xspace}
\def\Delm    {\ensuremath{\Delta m}\xspace}

\graphicspath{{./}}

\begin{document}

% \preprint{\babar-PUB-\BABARPubYear/\BABARPubNumber}
% \preprint{SLAC-PUB-\SLACPubNumber}

\begin{flushleft}
  \babar-PUB-\BABARPubYear/\BABARPubNumber \\
  SLAC-PUB-\SLACPubNumber \\
  hep-ex/\LANLNumber
\end{flushleft}

\begin{flushright}
\end{flushright}

\title{ \large \bf \boldmath
  Evidence of a Broad Structure at an Invariant Mass of 4.32~\gevcc 
  in the Reaction $\ee\to\pipi\psip$ Measured at \babar
}

%% author list as of 06-Jun-2006 (597 authors)
%
\author{B.~Aubert}
\author{R.~Barate}
\author{M.~Bona}
\author{D.~Boutigny}
\author{F.~Couderc}
\author{Y.~Karyotakis}
\author{J.~P.~Lees}
\author{V.~Poireau}
\author{V.~Tisserand}
\author{A.~Zghiche}
\affiliation{Laboratoire de Physique des Particules, F-74941 Annecy-le-Vieux, France }
% ...
\collaboration{The \babar\ Collaboration}
\noaffiliation

\date{\today}

\begin{abstract}

   We present a measurement of 
   the cross section of the process $\ee\to\pipi\psip$ 
   from threshold up to 8~\gev center-of-mass energy 
   using events containing initial-state radiation,
   produced at the \pep2~ $\ee$ storage rings. 
   The study is based on 298~\invfb of data recorded with the \BaBar 
   detector. A structure is observed in the cross-section not far above 
   threshold, near 4.32~\gev. 
   We also investigate the compatibility of this structure with 
   the \Ystat previously reported by this experiment.

\end{abstract}


\maketitle

Until recently, charmonium spectroscopy has been well described 
by potential models. 
Observations of the \Xstat~\cite{ct:X3872} 
and the \Ystat~\cite{ct:babar-Y} decaying into $\pipiJ$ 
complicate this picture, and 
have stimulated both experimental and theoretical interest in this area. 
The \Ystat can be produced by direct $\ee$ annihilation  
and is therefore known to have $J^{PC}=1^{--}$. 
Weak evidence for the \Ystat structure in $B$ decays 
was also reported by \BaBar~\cite{ct:babar-B-Y}. In addition, 
the \Ystat has been confirmed by the CLEO-c experiment 
in direct $\ee\to\Ystat$ interactions where the $\Ystat$ 
is detected in decays to $\pipi\jsi$ and $\ppz\jsi$~\cite{ct:cleo_c-Y_scan}; 
the observation of the latter mode and the measured ratio 
$\BF(\Ystat\!\to\ppz\jsi)/\BF(\Ystat\!\to\pipi\jsi)\approx0.5$ 
implies that the $\Ystat$ has isospin zero, 
as expected for a charmonium state.

It is peculiar that the \Ystat is wide and yet has a large
branching fraction into the hidden charm mode $\pipiJ$,
and that at the \Ystat mass the cross section for $\ee\to\text{hadrons}$ 
exhibits a local minimum.
Many theoretical interpretations for the \Ystat have been proposed, 
including unconventional scenarios:  
quark-antiquark gluon hybrids~\cite{ct:Y4260-hybrid} 
and hadronic molecules~\cite{ct:Y4260-molecule}. 
We undertook this study with the intent of clarifying the nature of the \Ystat.

In this Letter we study the process 
$\ee\to \pipi\psip$, $\psip\to\pipi\jsi$, 
for $\ee$ center-of-mass (CM) energies 
from threshold up to 8~\gev 
using initial-state radiation (ISR) events. 
The ISR cross section for a particular hadronic final state $f$ 
is given by
\begin{equation}\label{eq:ISR_XSect-dx}
     \frac{d\,\sigma_f(s,x)}{dx} = W(s,x)\cdot\sigma_f(s(1-x)),
\end{equation}
where $s$ is the square of the $\ee$ CM energy, 
$x\equiv 2E_{\gamma}/\sqrt{s}$ is the ratio of the photon energy 
to the beam energy in the $\ee$ CM frame, 
and $W(s,x)$ is the spectrum for ISR photon emission 
for which we use a calculation good to $\order(\alpha^2)$; 
the effective CM energy $\sqrt{s'}$ is 
the invariant mass of the final state $m=\sqrt{s(1-x)}$. 

We use data recorded with the \BaBar detector~\cite{ct:babar-detector} 
at the PEP-II asymmetric-energy $\ee$ storage rings, located at 
the Stanford Linear Accelerator Center.  
These data represent an integrated luminosity of 272~\invfb recorded 
at $\sqrt{s}$ = 10.58~\gev, near the $\Upsilon(4S)$ resonance, 
and 26~\invfb recorded near 10.54~\gev.

Charged-particle momenta are measured in a tracking system consisting 
of a five-layer double-sided silicon vertex tracker (SVT) and a 
40-layer central drift chamber (DCH), both situated in a 1.5-T axial 
magnetic field. 
An internally reflecting ring-imaging Cherenkov detector (DIRC) 
provides charged-particle identification. A CsI(Tl) electromagnetic 
calorimeter (EMC) is used to detect and identify photons and 
electrons, while muons are identified in the instrumented 
magnetic-flux return system (IFR).

Optimized selection criteria are chosen based on a simulated sample 
of $\ee\to\gISR\pipi\psip$ events and a sample of $\ee\to\gISR\psip$, 
$\psip\to\pipi\jsi$ candidates in data, which serves as a clean control 
sample~\cite{ct:lou-psi2s}.

A candidate $\jsi$ meson is reconstructed via its decay to $\ee$ or $\mm$. 
The lepton tracks must be well reconstructed, and at least one must be
identified as an electron or a muon candidate.
An algorithm to recover energy lost to 
bremsstrahlung is applied to electron candidates.
An $\ee$ pair with its invariant 
mass within the interval of (-100,+40)~\mevcc of the nominal $\jsi$ mass 
is taken as a $\jsi$ candidate. 
For a $\mm$ pair, the interval is (-60,+40)~\mevcc. 
The $\jsi$ candidate is then kinematically constrained to 
the nominal $\jsi$  mass
and combined with a pair of oppositely-charged tracks 
identified as pion candidates. 
The $\pi^+\pi^-\jsi$ combinations with invariant mass within 10~\mevcc 
of the nominal $\psip$ mass 
are taken as $\psip$ candidates. 
Another pair of oppositely-charged pion candidates (primary pions) 
is then combined with the $\psip$ candidate. 
The $\pipi\psip$ mass-resolution function is well described 
by a Cauchy distribution~\cite{ct:Cauchy} with a FWHM of about 7~\mevcc.
We do not require observation of the ISR photon ($\gISR$) 
as it is preferentially produced along the beam directions.

We select $\ee\to\gISR\pipi\psip$ events with the following  
criteria: 
(1) there must be no additional well-reconstructed charged tracks in the event;
(2) there must be no well-reconstructed $\pi^0$ 
    or $\eta\to\gamma\gamma$ in the event;
(3) the transverse component of the visible momentum in the $\ee$ CM frame, 
    including that of the $\gISR$ when it is reconstructed, 
    must be less than 2~\gevc;
(4) the difference ($\DelP$) between the measured $\pipi\psip$ momentum  
    and the value expected for it in an ISR $\pipi\psip$ event, that is, 
    $(s-m^2)/(2\,\sqrt{s})$, where $m$ is the $\pipi\psip$ invariant mass,
    must be within $[-0.10,+0.06]$~\gevc;
(5) $\cos\theta_\ell$,
    where $\theta_\ell$ is the angle between the lepton $\ell^+$
    momentum in the $\jsi$ rest frame and the 
    $\jsi$ momentum in the $\ee$ CM frame, 
    must satisfy $|\cos\theta_\ell| < 0.90$; and
(6) the invariant mass of the $\pipi$ pair in $\psip$ decay 
    must be greater than 0.4~\gevcc 
    in order to suppress the combinatorial $\pipi\jsi$ background.

A clean $\psip$ signal is apparent in Fig.~\ref{fg:bwFit-psi2S}. 
An examination of the $\pipi\psip$ combinations reveals that 
about half the background results from recombinations within the same 
$\fourPiJsi$ system where at least one of the primary pions is combined 
with the $\jsi$ to form a $\pipi\jsi$ candidate.  
After subtracting the self-combinatorial background, we estimate 3.8$\pm$1.1 
non-$\psip$ background events 
in the final sample of 78 events within the $\psip$ mass window.

\bfg[htbp]
 \centering
   \infg[width=8.8cm]{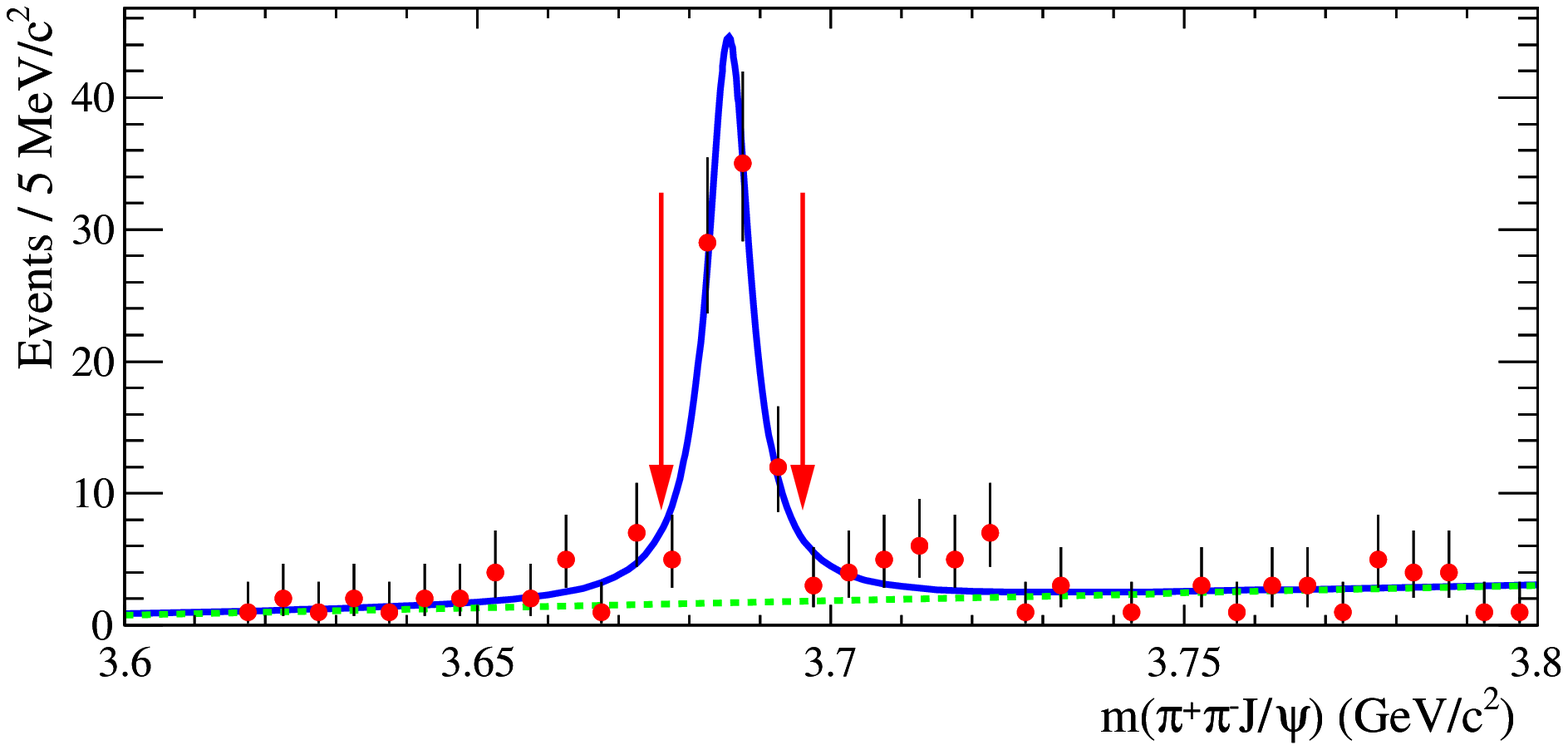}
\caption{
The invariant mass distribution for all $\pipi\jsi$ candidates 
where more than one entry per event is allowed. 
The solid curve is a fit to the distribution in which the $\psip$ signal 
is described by a Cauchy function and the background by a quadratic function 
(represented by the dashed curve). 
The arrows indicate the $\psip$ mass window.
}
\label{fg:bwFit-psi2S}
\efg

In Fig.~\ref{fg:isrVars} the  
distributions of (a) $\DelP$  
and (b) $\cos\theta^*$ for $\fourPiJsi$ candidates, 
where $\theta^*$ is the angle between the positron beam and the
$(\pipi\pipiJ)$ momentum in the $\ee$ CM frame,  
are shown and compared to expectations from simulations. 
There are 16 events that have a well reconstructed gamma 
with energy greater than 3~\gev, 
while the Monte Carlo simulation predicts 16.4 
for the same total number of ISR $\pipi\psip$ candidates. 
Furthermore, all events within $\abs{\cos\theta^*}<0.9$ 
are accompanied by 
a reconstructed gamma with energy greater than 3.0~\gev. 
We find excellent agreement in the ISR characteristics between the data 
and signal Monte Carlo sample. 
The good agreement in the $\DelP$ distribution rules out 
any significant feeddown from higher mass charmonia decaying 
to the $\psip$ with one or more undetected particles. 
As an example, the $\DelP$ distribution 
for $\psi(4415)\to\pipi\pi^0\psip$ events 
would peak around $-0.2\gevc$ with a long tail extending to 
well below $-0.2\gevc$.
We estimate the non-ISR $\pipi\psip$ background to be less than 1 event.

\bfg[htbp]
 \centering
   \infg[width=8.8cm]{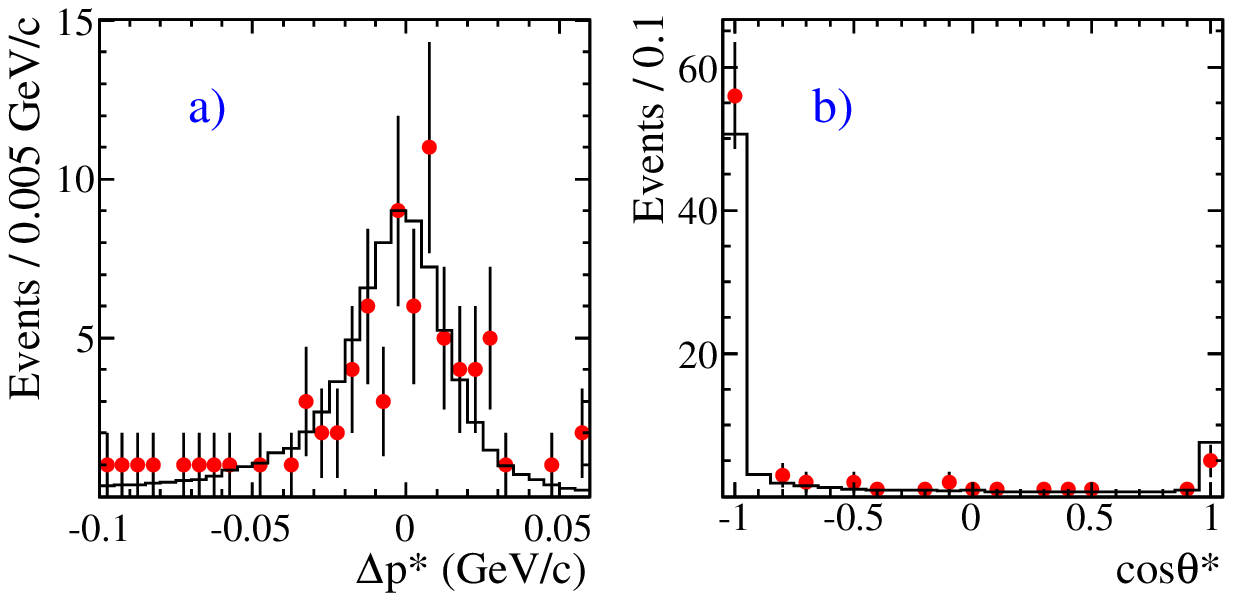}
\caption{
The distributions of 
(a) $\Delta{p^*}$ and (b) $\cos\theta^*$ of 
the $\fourPiJsi$ combination in the $\ee$ CM frame 
are shown for data (solid dots) and Monte Carlo simulation of the 
signal (histogram) normalized to the total number of the observed data events.
}
\label{fg:isrVars}
\efg

The track quality, particle identification information, and kinematic
variables of all pion candidates are examined, 
and displays of the events are scanned visually to
check for possible track duplications and other potential problems. 
No evidence for improper reconstruction or event quality problems is found. 

The $\fourPiJsi$ invariant-mass spectrum up to 5.7~\gevcc 
for the final sample is 
represented as data points in Fig.~\ref{fg:fitResult}.
A structure around 4.32~\gevcc is observed in the mass spectrum. 

To clarify the peaking structure observed in Fig.~\ref{fg:fitResult},  
we perform an unbinned maximum likelihood fit to the mass spectrum 
up to 5.7~\gevcc in terms of a single resonance 
with the following probability density function (PDF):
\begin{equation}\label{eq:sumPDF}
\begin{split}
  P(m) & = N\,a\cdot\eff(m) \cdot\Bigl(W(s,x)\cdot 2m/s\Bigr) 
           \cdot\frac{12\pi}{m^2} \\
       & \cdot \frac {M^2\cdot\Gamma_{ee}\cdot\Gamma_f\cdot \bigl(\Phi(m)/\Phi(M)\bigr)}
                     { (M^2-m^2)^2 + (M\,\Gamma_\text{tot})^2 }
                  \; + \; B(m)\;,
\end{split}
\end{equation}
where $M,\Gamma_\text{tot}, \Gamma_{ee}, \Gamma_f, N$ are 
the nominal mass, total width, partial width to $\ee$, 
partial width to 
$\pipi\psip$, and yield for a resonance, respectively, 
and $m$ is the $\fourPiJsi$ invariant mass,
$\eff(m)$ is the  mass-dependent efficiency,
$\Phi(m)$ is the  mass-dependent 
phase-space factor for a $S$-wave three-body $\pipi\psip$ system,
$a$ is a normalization factor,
and $B(m)$ is the PDF 
(the shaded histogram in Fig.~\ref{fg:fitResult})
for the non-$\psip$ background. 
The shape of $B$ was obtained from $\psip$ sideband events 
with its integral fixed to 3.1 events corresponding to 
the mass region in the fit, 
where the total number of events is 68.
The mass dependence of $\Gamma_\text{tot}$ 
is ignored in the fit. 

\bfg[htbp]
 \centering
   \infg[width=8.8cm]{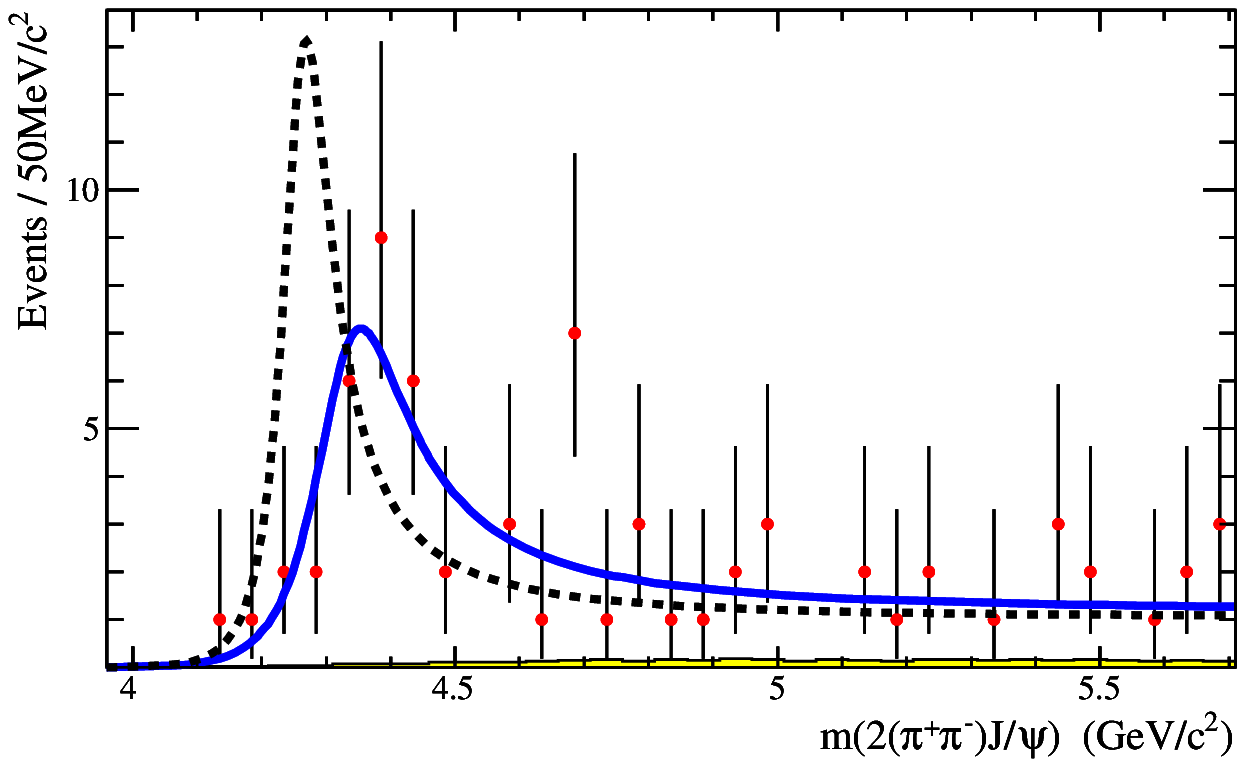}
\caption{
  The $\fourPiJsi$ invariant mass spectrum up to 5.7~\gevcc 
  for the final sample. 
  The shaded histogram represents the fixed background and
  the curves represent the fits to the data (see text).
}
\label{fg:fitResult}
\efg

%%%%%
We perform fits to the distribution in Fig.~\ref{fg:fitResult} 
to test hypotheses that 
the data are a result of the decay of the \Ystat (dashed curve) using 
resonance parameters fixed to those of Ref~\cite{ct:babar-Y}, 
and alternatively those of the $\psi(4415)$ (not shown) 
with the mass and width taken from Ref~\cite{ct:PDG2006}. 
In the third fit (solid curve) we assume a single resonance 
whose mass and width are free parameters, which are then found to 
be $(4324\pm24)~\mevcc$ and 
$(172\pm33)~\mev$ (after unfolding mass-resolution) by the fit. 
We calculate the $\chi^2$\!/dof value for each fit to test these hypotheses. 
In the calculation the events in Fig.~\ref{fg:fitResult} are 
regrouped so that at least seven events are expected in each bin. 
The $\chi^2$\!/dof values are found to be $21.3/8$, $54.4/7$, and $7.3/7$ 
for hypotheses of the \Ystat, the $\psi(4415)$, and a new resonance, 
respectively, 
corresponding to $\chi^2$-probabilities of $6.5\times10^{-3}$, 
$2.0\times10^{-9}$, and $29\%$. 
The low probabilities associated with the \Ystat and the $\psi(4415)$ 
indicate that the structure is not consistent with the $\psi(4415)$, 
and is not well described by the \Ystat either.
We also perform a fit including both the \Ystat and $\psi(4415)$ 
plus their interference, and find the $\chi^2$\!/dof value to 
be $17.8/6$, corresponding to a $\chi^2$-probability of $6.7\times10^{-3}$, 
but no much improvement from the fit to the \Ystat only. 
 In order to further compare the structure reported here with 
 the \Ystat reported in Ref.~\cite{ct:babar-Y}, 
 we perform simultaneous fits to
 both the $\pipi\psip$ mass spectrum in Fig.~\ref{fg:fitResult} 
 and the $\pipi\jsi$ mass distribution in \cite{ct:babar-Y}
 under the hypotheses that 
   (1) both signals are a single resonance 
 and 
  (2) these signals are manifestations of two independent resonances, 
with a single resonance for each signal. 
The PDF as used in Ref.~\cite{ct:babar-Y} is applied to the fit to 
the $\pipi\jsi$ mass distribution. 
 The logarithmic likelihood obtained 
 from  the single-resonance hypothesis (1) 
 is 5.4 units less than that obtained from 
 the two-resonance hypothesis (2), which 
 corresponds to a $\chi^2$-probability of $4.5\times10^{-3}$ 
 for the single-resonance hypothesis assuming 
 a $\chi^2$ distribution for  
 the difference in the logarithmic likelihood between 
 the two hypotheses. 
 However, none of the probabilities associated with the \Ystat 
 can exclude the possibility that the structure observed is 
 a manifestation of a new decay mode for the \Ystat.

%%%%%

The primary $\pipi$ invariant mass distribution for the selected events 
within $m(\fourPiJsi)<5.7~\gevcc$ is shown in Fig.~\ref{fg:m2pi}. 
For the two events having more than one $\psip$ candidates, 
the dipion invariant mass is only included for the $\psip$ candidate 
closest to its nominal mass. The Monte Carlo distribution is also shown 
in Fig.~\ref{fg:m2pi} for a single resonance decaying to $\pipi\psip$ 
in a $S$-wave three-body phase-space using the resonance parameters 
obtained in the above paragraph.

\bfg[htbp]
 \centering
   \infg[width=8.8cm]{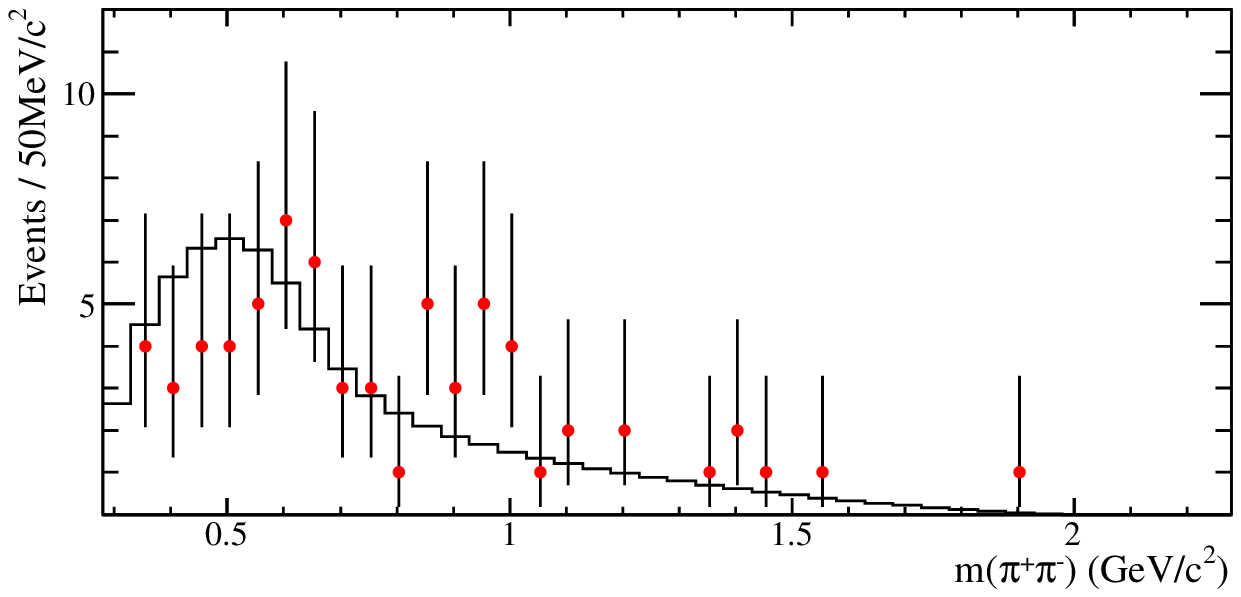}
\caption{
  The primary $\pipi$ invariant-mass spectrum 
  within region $m(\fourPiJsi)<5.7~\gevcc$ for the final sample. 
  Only one entry per event is included in the plot, 
  as described in the text. 
  The histogram shows the distribution for Monte Carlo events 
  (see text).
}
\label{fg:m2pi}
\efg

We extract the energy-dependent cross section for 
$\ee\to\pipi\psip$ up to 8~\gev for the final sample. 
The average cross section over a mass range of width $\Delm$ 
is calculated as 
\begin{equation}\label{eq:XSect-average}
\begin{split}
    \overline{\sigma(m)} 
    & \equiv \int_{m-\Delm/2}^{m+\Delm/2} \sigma(x)\,dx \,\big{/}\,\Delm \\ 
    & \approx \frac{1}{\Lumi\cdot\BF\cdot\Delm}
    \sum_{i} \Bigl(  \frac{1} {2m_i/s\cdot W(s,1-m_i^2/s)\cdot \eff_i} \Bigr)\;,
\end{split}
\end{equation} \\
where $\Lumi$ is the integrated luminosity,
$\BF$ is the product of $\BF(\psip\to\pipi\jsi)$ and $\BF(\jsi\to\ll)$,
the sum is over all events within the mass range,
$m_i$ is the $\fourPiJsi$ invariant mass, 
and $\eff_i$ is the estimated efficiency at that mass. 
The measured cross section is shown in Fig.~\ref{fg:XSect} 
and the numerical results can be found in~\cite{ct:XSect-table}, 
where the background has been subtracted from bins with non-zero content. 
The energy-dependent selection efficiency 
(solid histogram in Fig.~\ref{fg:XSect}) is determined 
from Monte Carlo events for which the $\psip$ polarization has been 
properly considered while the primary $\pipi$ is generated 
in $S$-wave phase-space. The uncertainty in the selection efficiency 
due to model dependence is estimated from the efficiency difference 
between $S$-wave phase-space model and multipole model~\cite{ct:VVPIPI} 
in the primary $\pipi$ generation.
The main systematic uncertainties are listed in Table~\ref{tb:sysError}, 
and are added in quadrature, resulting in 
a total systematic uncertainty of $12.3\%$. 

\bfg[htbp]
 \centering
   \infg[width=8.8cm]{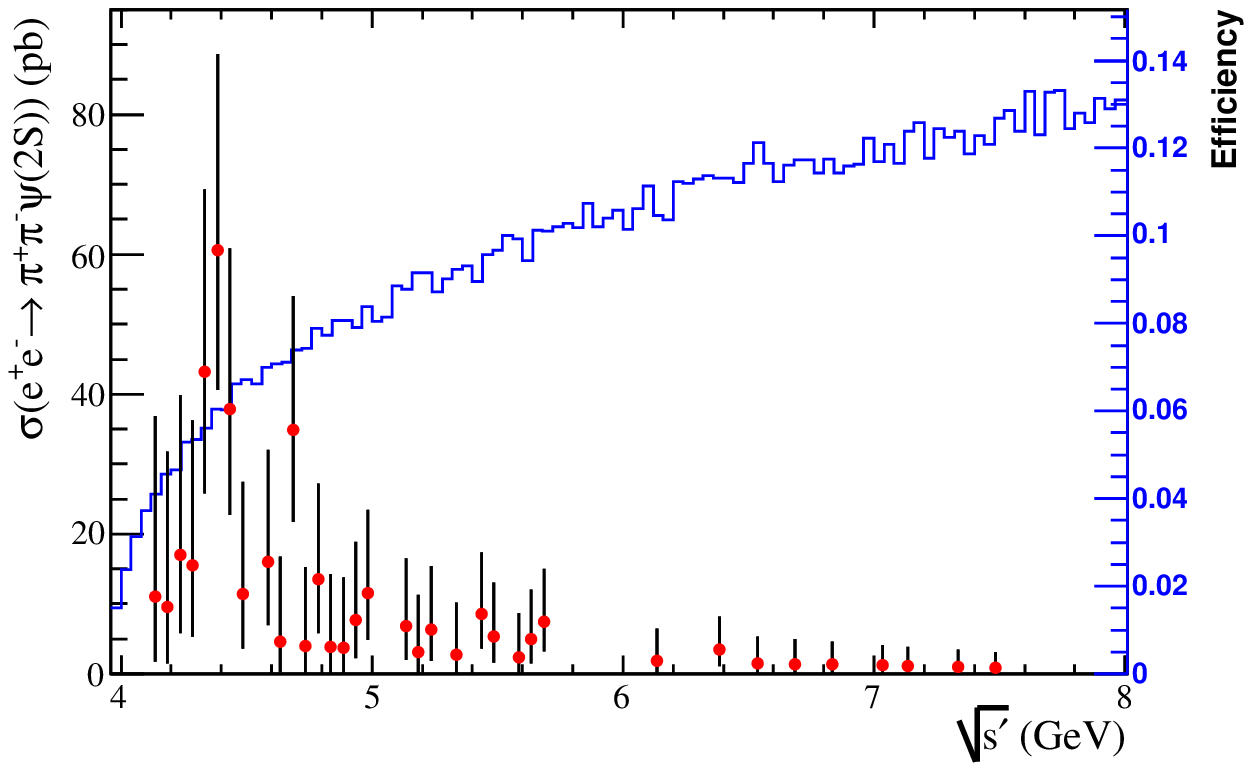}
\caption{
   The measured CM energy dependence of the cross section 
   (points with error bars)
   for $\ee\to\pipi\psip$ after background subtraction. 
   The solid histogram shows the energy-dependent selection efficiency.
}
\label{fg:XSect}
\efg

\btbl[hbtp]
\caption{Summary of main systematic uncertainties for 
         the $\ee\to\pipi\psip$ cross section measurements.
        }
\label{tb:sysError}
\begin{ruledtabular}
\btbu{c c}
     Source                      & Systematic error
\\ \hline
    Model-dependent acceptance   & $\pm9.0\%$
\\
   Tracking efficiency           & $\pm7.6\%$
\\
   $\BF(\psip\to\pipi\jsi)\cdot\BF(\jsi\to\ll)$  & $\pm3.5\%$
\\ \hline
   Total                         & $\pm12.3\%$
\etbu
\end{ruledtabular}
\etbl

In summary, we have used ISR events to study the exclusive process
$\ee\to\pipi\psip$ and to measure its energy-dependent cross section 
from threshold to 8~\gev CM energy. 
A structure is observed at $\sim4.32~\gevcc$ 
in the $\pipi\psip$ invariant mass spectrum that is 
not consistent with the decay $\psi(4415)\to\pipi\psip$. 
A fit to the mass spectrum with a single resonance yields 
a mass of $(4324\pm24)~\mevcc$ and a width of $(172\pm33)~\mev$, 
where the errors are statistical only. 
The structure in Fig.~\ref{fg:fitResult} has a mass that differs somewhat 
from that reported for the \Ystat in Ref.~\cite{ct:babar-Y}.  
However, the possibility that it represents evidence for a new decay mode 
for the \Ystat cannot be ruled out at this time.

We are grateful for the excellent luminosity and machine conditions
provided by our \pep2\ colleagues, 
and for the substantial dedicated effort from
the computing organizations that support \babar.
The collaborating institutions wish to thank 
SLAC for its support and kind hospitality. 
This work is supported by
DOE
and NSF (USA),
NSERC (Canada),
IHEP (China),
CEA and
CNRS-IN2P3
(France),
BMBF and DFG
(Germany),
INFN (Italy),
FOM (The Netherlands),
NFR (Norway),
MIST (Russia), and
PPARC (United Kingdom). 
Individuals have received support from the
Marie Curie EIF (European Union) and
the A.~P.~Sloan Foundation.

\end{document}